\documentclass{article}

\usepackage[preprint]{neurips_2023}

\usepackage[utf8]{inputenc} 
\usepackage[T1]{fontenc}    
\usepackage{hyperref}       
\usepackage{url}            
\usepackage{booktabs}       
\usepackage{amsfonts}       
\usepackage{nicefrac}       
\usepackage{microtype}      
\usepackage{xcolor}         

\title{Decoding The Digital Fukú: Deciphering Colonial Legacies to Critically Assess ChatGPT in Dominican Education}

\author{%
Anaelia Altagracia Ovalle \\
  Department of Computer Science\\
  Universiy of California, Los Angeles\\
  \texttt{anaelia@cs.ucla.edu}
}

\begin{document}

\maketitle

\begin{abstract}

Educational disparities within the Dominican Republic (DR) have long-standing origins rooted in economic, political, and social inequity.\footnote{https://www.education-inequalities.org/countries/dominican-rep} Addressing these challenges has necessarily called for capacity building with respect to educational materials, high-quality instruction, and structural resourcing. Generative AI tools like ChatGPT have begun to pique the interest of Dominican educators due to their perceived potential to bridge these educational gaps. However, a substantial body of AI fairness literature has documented ways AI disproportionately reinforces power dynamics reflective of jurisdictions driving AI development and deployment policies, collectively termed the \textit{AI Global North}. As such, indiscriminate adoption of this technology for DR education, even in part, risks perpetuating forms of digital coloniality. Therefore, this paper centers embracing AI-facilitated educational reform by critically examining how AI-driven tools like ChatGPT in DR education may replicate facets of digital colonialism. We provide a concise overview of 20th-century Dominican education reforms following the 1916 US occupation. Then, we employ identified neocolonial aspects historically shaping Dominican education to interrogate the perceived advantages of ChatGPT for contemporary Dominican education, as outlined by a Dominican scholar. This work invites AI Global North \& South developers, stakeholders, and Dominican leaders alike to exercise a relational contextualization of data-centric epistemologies like ChatGPT to reap its transformative benefits while remaining vigilant of safeguarding Dominican digital sovereignty.

\end{abstract}

\section{Introduction}

In Dominican lore, \textit{fukú} describes the curse brought to Hispañola on slave ships by Spanish imperialists in their pursuit of the New World. It is a persistent doom casted upon the island and its inhabitants since Colonization, reverberating throughout generations as manifestations of political, economic, social, and personal misery [\citenum{FerbelAzcarate2002NotEW}, \citenum{diaz_brief_2008}, \citenum{mcconnell2013black}, \citenum{moyer_classicisms_2020}]. We ground this paper in the colonial imaginary to center ways  interlocking forms of coloniality have endured across dimensions of Dominican life. The educational system in the Dominican Republic (DR) is no exception, historically grappling with instability and inequality stemming from historical colonial influences that persist to this day. Colonial forces exhibit a fluid nature, manifestating in the digital contemporary as digital coloniality [\citenum{Ricaurte2019DataET}]. This is marked by the prevalence of data-centric epistemologies and technologies which entrench power asymmetries [\citenum{Madianou2019TechnocolonialismDI}], leading to critical AI fairness researchers raising concerns on the potential for epistemic violence in AI-driven technologies like ChatGPT. These technologies often exhibit a bias toward predominantly white and Anglo-Saxon perspectives, favoring AI global north contexts while marginalizing underrepresented voices [\citenum{mohamed2020decolonial, ovalle2023factoring, Bianchi2022EasilyAT}]. Contextualized to the Dominican Republic's educational system, we identify parallels between neocolonialism and present-day techno-optimism. Responding to existing work by Dominican scholars who share enthusiasm for incorporating technologies like ChatGPT into Dominican Education, this paper aims to assess risks in manifesting digital \textit{fukú} through the inadvertent relinquishing of Dominican intellectual sovereignty via the uncritical adoption of these technologies. While ChatGPT may contribute to expanded educational access, we strongly urge Dominican public institutions like  the Ministry of Education, AI developers from the Global South, and, most importantly, Dominican citizens from various socioeconomic backgrounds, to actively participate in a critical assessment of these technologies within the context of their colonial histories. This approach will enable a deeper understanding of the sociotechnical implications they carry, so as to empower stakeholders to make more informed decisions surrounding their digital agency.

\vspace{-0.3cm}
\section{Divisions between The AI Global North and AI Global South}

The terms \textit{AI Global North} and \textit{AI Global South} are often used to describe the disparities in the development, deployment, and impact of artificial intelligence technologies between regions of the world [\citenum{png2022tensions}]. The predominant dissemination of AI Global North-oriented technology and AI governance reflect resource inequalities, infrastructural gaps, and varying concentrations of power [\citenum{Tang2022ASN, muro-liu-2021, Solaiman2023EvaluatingTS}]. This concentration of authority reinforces dominant values [\citenum{mohamed2020decolonial}] onto vulnerable communities, thereby exacerbating inequality and harm [\citenum{Solaiman2023EvaluatingTS}]. This results in the creation of LLMs whose sociolinguistic boundaries reflect formations of privilege and power [\citenum{Rickford1985EthnicityAA, rosa_flores_2017}]. As just one example, the most commonly used large language models (LLM) are pretrained on data primarily sourced from English, anglo-centric texts located in North America and the UK [\citenum{gururangan2022whose}]. As a result, less resourced languages are more prone to jailbreaking an LLM [\citenum{yong2023low}] and non-native English speakers are more likely to be flagged for plagiarism by an AI system [\citenum{liang2023gpt}].  

This \textit{digital privileging} is reminiscent of colonial legacies which wiped indigenous knowledge through coercive cultural assimilation and exploitative practice. Generative AI systems derived by the Global North are developed and trained within AI Global North contexts, holding societal impacts tied to propagating stereotypes, representational and allocation harm, and cultural erasure [\citenum{Solaiman2023EvaluatingTS}]. Empirical risk minimization (ERM), a theoretical foundation underpinning many machine learning (ML) algorithms, is a general method for fitting some parametrized predictor (e.g. linear regression, logistic regression, neural networks) on real-world data. ERM finds model parameters that minimize a user-defined loss function by taking its expectation, resulting in an \textit{averaging} of the loss over a training dataset. Yet, it is through this expectation that majority groups are privileged in representation learning, as minimizing average error fits majority populations [\citenum{chouldechova2018frontiers, bell2023simplicity}]. This subsequent tyranny of the majority [\citenum{Chouldechova2018TheFO}] leads language representations to favor the majority groups. English, American culture and lexicon [\citenum{gururangan2022whose}] are the prevailing data used for LLM pretraining. Consequently, these behaviors lead to direct and indirect harms on non-dominant groups, inclusive of vulnerable and marginalized communities.  It is through this severe power asymmetry that we recognize parallels to coloniality that endanger a country's structural fabrics within AI Global South, including a country's educational system. Notably, this may occur even in the absence of malicious intent, even when individuals act in good faith.


In practice, existing inequities are algorithmically exacerbated in many forms. AI detectors consistently misclassify non-native English writing samples as AI-generated [\citenum{liang2023gpt}]. Biases appear against minoritized groups, where LLMs have displayed, for instance anti-Muslim [\citenum{Abid2021PersistentAB}] and anti-LGBTQIA behavior [\citenum{queerinai2023queer}]. [\citenum{xu2021detoxifying}, \citenum{santy2023nlpositionality}] found that LLMs overfilter voices from historically marginalized communities during LLM detoxicifcation. Meanwhile, users are increasingly using LLMs to analyze information, even though these models have shown persistent inclinations to harms and hallucination with respect to factual information [\citenum{Ji2022SurveyOH}, \citenum{Zhou2020DetectingHC}]. In the absence of diverse social and infrastructural investment defending against an undue sense of trust and dependence on these systems [\citenum{Solaiman2023EvaluatingTS}], dissemination of biased and inaccurate information have large scale ramifications. Therefore, ensuring the trustworthiness of AI necessarily calls for their careful situating within AI Global South contexts. While several works have examined aspects of coloniality in AI, none have contextualized what this may mean for the Dominican Republic. In the following section, we provide a concise historical overview of education in the DR, which later helps contextualize some implications of incorporating ChatGPT into the DR educational system.

\vspace{-0.3cm}
\section{Dominican Republic: Historical Grapplings with Education and Agency}

The Dominican Republic has consistently grappled with domination by external entities. \footnote{Initially colonized by Spain, the country gained independence in 1821, briefly occupied by Haiti from 1822 to 1844, until going under Spanish rule again until 1865.}. At the turn of the 20th century, the DR faced US occupation between 1916-1924. During this time, the DR's education system went through several overhauls. In this section, while limited by space in this paper, we do our best to identify a handful of threads which cogently unveil how 20th century neocolonialism shaped education in the Dominican Republic. We acknowledge that this recounting is not exhaustive.

\noindent \textbf{U.S. Paternalism \& White Supremacy} \indent At the turn of the 20th century, U.S. concerns of geopolitical instability \footnote{US investments in Dominican agriculture,  the Panama Canal} were heightened by the DR's debt to European creditors and the looming First World War. In addition to the US consolidating Dominican loans in exchange for control over DR customs, President Woodrow Wilson authorized the 1916 military occupation as a preemptive measure to protect U.S. assets, subsequently cementing the paternalistic role of the U.S. [\citenum{rodriguez2021prosperity}]. While framing their actions as altruistic, several historical documents authored by US military governors outline clear initiatives aimed at dissolving Dominican autonomy and identity. Working-class adult Dominicans were described as ``mulatto peasants'' and children which lacked mental capacity for self-sovereignty, reflecting clear forms of explicit racial minoritization [\citenum{USMC_Handbook, QuarterlyReport1918, PsychologicSituation1921, rodriguez2021prosperity}]. As such, overhauling the DR educational system was seen as a priority so that individuals could better participate as citizens [\citenum{rodriguez2021prosperity}]. Simultaneously, Dominican intellectual elites, termed \textit{letrados}, varied in their opinion of the Dominican \textit{campesinos}, or rural masses. Some embraced their Black lineage and \textit{nueva raza} while others fed white supremacist ideology, characterizing lower social classes --  often non-white -- as ``semi-savage'', apathetic, and indifferent to their duties as citizens [\citenum{MartinezVergne2006, ubiera2015contrapunteo, rodriguez2021prosperity}]. These \textit{letrados}, in charge of instilling US mandates post-occupation, were the same citizens advocating for increased immigration of Europeans so as to dilute the DR's African heritage [\citenum{calder1984impact, MartinezVergne2006, Derby2009TheDS}]. 

\noindent \textbf{Coercive Tutelage} \indent
Notably, the DR iterated on educational reform prior to US occupation, with its ``Código Orgánico y Reglamentario de Educación Común'' issued in 1914 [\citenum{billini1999formacion}]. However, US officials introduced new reforms and a new educational code that centered American values and ensured alignment with US interests. This was enacted through the appointing of American allies in key government positions and implementing educational reforms. Reminiscent of Rudyard Kipling's White Man's Burden [\citenum{Kipling1899}], the US undertook efforts to ``civilize'' the nation through its teaching of democracy, citizenship, and capitalism [\citenum{rodriguez2021prosperity}]. Between 1917 and 1924, the US addressed high illiteracy rates in a coercive manner that further cemented social inequities. 
Redirecting funds from secondary schools to rural primary school construction, US officials solidified a labor division in the Dominican Republic; education prioritized preparing future agricultural laborers, thereby perpetuating existing social mobility gaps.
The US introduced enrollment mandates that obligated educators in the Dominican to police attendance and penalize parents and guardians of school children, so as to fulfill these mandates [\citenum{Ros2020ElIN, Gonzlez2018RecensinDS, nunez2022dominican, MartinezVergne2006, ZellerThesis}].

\noindent \textbf{Unsustainable Growth} \indent Rural schools grew faster than their ability to be well-resourced. Letters between Dominican educational administration evidence how they relied on the free labor from community members for rudimentary school construction. Even then, schools had large student-to-teacher ratios, with teachers paid up to 75\% less than others who taught in grade schools [\citenum{OrtegaFrier1920, rodriguez2021prosperity}]. In an attempt to improve resourcing, U.S. military government imposed property taxes, though this had limited benefit and only strained rural communities. As a result, the military government permanently closed many schools and centers for educator training, leaving Dominicans with gutted educational structures.

\vspace{-0.3cm}
\section{Assessing the Colonial Threads of ChatGPT for Dominican Education}

In an analysis of ChatGPT for Domininan education, \cite{bueno2023analysis} reminds readers that capacity building is long overdue in the Dominican Republic, optimistically framing the tool as a new avenue to educational resourcing and outlines benefits such as 1) access to educational material 2) personalized tutoring and 3) training instructors. Acknowledging ChatGPT's potential to influence pedagogy and educational access in the Dominican Republic, assessing its actual benefits for students and educators relies on a thorough evaluation within its existing education system. In this section, we position each claimed advantage within the crossroads of DR's neocolonial grapplings and digital coloniality. By demonstrating how AI tools can perpetuate these historical harms, our paper emphasizes that validating their benefits relies on empowering DR constituents to critically assess these tools.

\noindent \textbf{Access to Educational Material} \cite{bueno2023analysis} highlights ChatGPT's potential to widen access to educational materials otherwise previously inaccessible. While ChatGPT holds potential for providing educational materials on-the-fly, critical works in natural language processing (NLP) reveal the tendency of LLMs to hallucinate, potentially leading to the generation of incorrect or fake content [\citenum{Ji2022SurveyOH}]. Secondly, these works show significant deficiencies in LLM abilities to retrieve facts in non-English languages [\citenum{schott2023polyglot, Jiao2023IsCA}]. Therefore, non-English speakers engaging with the tool are more likely to consume low-quality or misleading information. Furthermore, access to ``educational material'' presupposes that text outputted by ChatGPT is universally relevant across cultural contexts. However, knowledge is not neutral; every LLM operates within specific frame of reference [\citenum{Solaiman2023EvaluatingTS}]. As such, the success of an LLM like ChatGPT hinges on the quality and diversity of its pretraining data so that it can operate across contexts. And yet, LLMs like ChatGPT predominantly learn from texts from or sponsored by the AI global north (e.g. preference annotations) [\citenum{ovalle2023factoring, gururangan2022whose}]. Fraught with familiar epistemic violence through the erasure of indigenous voices, LLMs which train on these hegemonies output ``material'' which can perpetuate power imbalances through the privileging of AI Global North values, norms, discourse, and knowledge which need to be scrutinized. Therefore, in assessing ChatGPT's viability for DR education, the ``view from nowhere'' must be rejected through the sociohistorical contextualization of these technologies.

\noindent \textbf{Personalized Digital Tutelage} \indent  
\cite{bueno2023analysis} points to ChatGPT as a new resource for assisting students in exam preparation, delivering real-time homework feedback and enabling personalized learning. While the technology holds promise, concerns linger in its impact on student learning and understanding student progress. For instance, potential plagiarism resulting from the LLM's generated content necessitate new approaches to fairly assessing student knowledge [\citenum{lo2023impact, cotton2023chatting}]. However, failing to incorporate critical sociocultural context in these approaches, especially if they incorporate AI tools, risks harming students. Namely, LLMs used for plagiarism detection exhibit biases against non-native English writers, resulting in their marginalization in both evaluative and educational settings \citenum{liang2023gpt}. This highlights yet another aspect of harm caused by Anglo-centric LLMs within multilingual contexts [\citenum{Gondwe2023CHATGPTAT, Jiao2023IsCA, rosa_flores_2017}]. English's global dominance stems from a historical process of militarized colonization, raising concerns about the erosion of multilingualism, language and culture rights, and perpetuation of marginalization through generative AI systems [\citenum{Solaiman2023EvaluatingTS}]. Meanwhile, scholarship which explores the benefits of ChatGPT-based tutoring remain within a western context [\citenum{lo2023impact}], which in turn, raise further concern over its ability to recognize alternative worldviews and historical accounts. Risks of imposing particular forms of historical knowledge by AI tools are reminiscent of the DR's grapple to maintain a united sense of identity and agency [\citenum{Siles2023FluidAI}]. Given aforementioned dependence on diverse training data, one may reasonably question the extent to which these tools can offer tutoring contextualized to varying sociohistorical positionalities, inclusive of those from the margins. Parallels nostalgically emerge between US military government in DR celebrating the expansion of schools as their own achievement, while community narratives offer a differing viewpoint which centers achieving this through local community efforts [\citenum{Rodriguez2022ANF}]. Within techno-optimistic discourse, it is critical to approach new AI tools with reasonable skepticism, especially if students may be privy to heavily rely on digital tutoring. Therefore, the prospect of AI tool adoption in DR education requires a careful examination of their contextualized tutoring benefits, the imparting of critical AI assessment skills to students, and new methods to fairly assess student knowledge  [\citenum{lo2023impact}].


\noindent \textbf{Teacher Aids and Training} \indent Another claim by \cite{bueno2023analysis} centers how ChatGPT can assist teachers in administrative work and professional training. LLMs like ChatGPT have the potential to revolutionize these aspects, though scholarly discourse on computer science and education describe the success of AI education as closely dependent on the readiness of teachers and their trust in AI tools [\citenum{ayanwale2022teachers, choi2023influence}]. Situated in a historical context, DR educational capacity building has faced substantial demand. Meeting this demand necessitates a comprehensive understanding of educators' needs, capabilities, and constraints, shaping case-specific guidance [\citenum{kasneci2023chatgpt}]. Open educational resources - like tutorials, studies, and guidelines -  offer a way for educators and institutions to gain knowledge about using LLMs in education  [\citenum{kasneci2023chatgpt}]. For instance, learning to differentiate between model-generated and student-generated answers is a valuable skill in AI-assisted pedagogical practice. Nevertheless, exclusively providing this training is no substitute for essential educator needs, such as fair teacher salaries and addressing persistent structural inequalities in educational funding and maintenance.

\noindent \textbf{Safety} \indent \cite{bueno2023analysis} emphasizes the need to secure the storage and protection of student personal information while using ChatGPT, raising concerns about third-party access. However, ChatGPT already owns user-submitted queries, thereby necessitating a reconceptualization of privacy and safety within this context. Reassessing safety implications within the existing Dominican educational system  presupposes clear definitions of digital privacy, agency, and safety infringements [\citenum{ovalle2023chatgpt}]. In this regard, gaps in digital literacy also hold significant risk to both the digital sovereignty of the user and their wider community. Previously mentioned scholarship detailing AI hallucinations and possibilities for consuming and propagating misinformation raise concerning questions around safety within and across these resolutions. If relying on ChatGPT as a sole educational resource, rather than an auxiliary resource, this concern for safety only amplifies [\citenum{Leite2023DetectingMW}]. Addressing these safety concerns requires building capacity which critically drives transparent discourse surrounding digital sovereignty  to both educators and parents alike. One approach may include dedicating resources for educational seminars and audits by educators, educational staff, and students so that they learn about data privacy, regulations, ethical concerns and best practices surrounding data protection [\citenum{kasneci2023chatgpt}]. Furthermore, despite being presented as a transformative tool free of charge, dimensions of invisible data labor by users paradoxically reify the notion that user data - the driving force behind machine learning - is exclusively created by corporations. Consequently, this reinforces the idea that corporations should be the sole custodians of such data. For instance, eliciting user preferences with a thumbs up or thumbs down after ChatGPT generations is one data point consumed for RLHF training [\citenum{arrieta2018should}, \citenum{li2023dimensions}, \citenum{casper2023open}]. While one may argue that free user-tailored systems already brings substantial benefits to users, masking this data labor prevents broader public discourse concerning the equitable distribution of gains within the data economy. In unveiling these aspects, addressing safety concerns goes beyond merely acknowledging the need to secure personal information. It necessitates both questioning and understanding the safety implications behind (1) \textit{who} is responsible for safeguarding personal data (2) \textit{how} is that data used, and (3)  
our capacity to recognize potential misinformation and manage unsafe or adverse AI-driven outcomes.

\vspace{-0.3cm}
\section{Conclusion}
The Dominican Republic's colonial history reveals important historical considerations for assessing ChatGPT for Dominican education. While the prospect of integrating the AI-driven tool for improved educational access drives speculative optimism, validating its benefits on-the-ground relies on DR constituents holistically assessing these tools for themselves. This requires centering the exercising of digital agency in praxis; empowering Dominican citizens to critically assess AI-driven systems requires digital capacity building that is supported by structural initiatives aimed at broadening access to AI technologies in a way reflects the will of its constituents.

\bibliographystyle{plainnat}
\bibliography{custom.bib}

\end{document}